# Direct Evidence of Mg Incorporation Pathway in Vapor-Liquid-Solid Grown *p*-type Nonpolar GaN Nanowires

*Avinash Patsha,[1,*] S. Amirthapandian,[2] Ramanathaswamy Pandian,[1] S. Bera,[3,*] Anirban Bhattacharya,[4] and Sandip Dhara[1,*]*

[1]Surface and Nanoscience Division, Indira Gandhi Center for Atomic Research, Kalpakkam-603102, India

[2]Materials Physics Division, Indira Gandhi Center for Atomic Research, Kalpakkam-603102, India.

[3] Water and Steam Chemistry Laboratory, BARC Facilities, Kalpakkam-603102, India

[4] Institute of Radiophysics and Electronics, University of Calcutta, Kolkata-700009, India

**Corresponding Authors:** avinash.phy@gmail.com; bera@igcar.gov.in, dhara@igcar.gov.in

*Abstract*

Doping of III-nitride based compound semiconductor nanowires is still a challenging issue to have a control over the dopant distribution in precise locations of the nanowire optoelectronic devices. Knowledge of the dopant incorporation and its pathways in nanowires for such devices is limited by the growth methods. We report the direct evidence of incorporation pathway for Mg dopants in *p*-type nonpolar GaN nanowires grown via vapour-liquid-solid (VLS) method in a chemical vapour deposition technique for the first time. Mg incorporation is confirmed using X-ray photoelectron (XPS) and electron energy loss spectroscopic (EELS) measurements. Energy filtered transmission electron microscopic (EFTEM) studies are used for finding the Mg incorporation pathway in the GaN nanowire. Photoluminescence studies on Mg doped GaN nanowires along with the electrical characterization on heterojunction formed between nanowires and *n*-Si confirm the activation of Mg atoms as *p*-type dopants in nonpolar GaN nanowires.

**KEYWORDS:** Mg doped GaN, nonpolar GaN, dopant incorporation pathway, heterostructure *p-n* junction, photoluminescence

The functionality of nanoscale electronic and optoelectronic devices made of compound semiconductor nanowires like group III nitrides rely on the complex process of doping. The efforts put on the incorporation of dopants in the III nitride based GaN nanowires leads to realise nanodevices of LED, lasers, high electron mobility transistor (HEMT), logic gates, photodetectors, solar cells and gas sensors.[1-7] Apart from the doping process the type of dopant and defects, variation in nanowires size and crystallographic orientations heavily influence the growth and various electrical and optical properties of nanowires.[8-12] Generally the nanowires are grown via catalyst mediated vapor-liquid-solid (VLS) method which offers the control over diameter and composition of the nanowires.[9,13] For the growth of compound semiconductor like III nitride nanowires in VLS process, the liquid/solid interface involving the catalyst and the nanowire plays a crucial role in driving axial growth of the nanowire. Any abrupt variations in composition at the interface deviates the nanowire growth from steady state.

Mg is the most successful *p*-dopant in III-nitrides,[14-16] which pave the way for realising III-nitride based electronic,[15,16] and optoelectronic devices.[3,4,17] Integration of GaN with Si has been carried out by forming a vertically aligned *p*-GaN nanorods/*n*-Si heterojunction in the Au catalyst assisted halide chemical vapour deposition (CVD) technique and demonstrated as photovoltaic cell.[3] Doping of III nitride nanowires, in particular *p*-type doping is a difficult process compared to that for thin films as the dopant incorporation and distribution depends on N/III ratio and incorporation pathways, respectively.[18,19] Depending on the dopant type, host lattice and crystallographic orientation the incorporation pathways can alter in nanowires as well as in thin films.[20-22] Understanding of the dopant incorporation paths in nanowires can help in tuning the dopant distribution over a precise location along the nanowire which is crucial for aforementioned nanowire devices. Mg incorporates in GaN by substituting in place of Ga atom.

Because of its high-acceptor ionization energy in GaN, however concentration of unionized Mg dopants are high at room temperature leading to low hole density. Also the native point defects like nitrogen vacancies ($V_N$) and unintentional impurity (oxygen) atoms acts as donors which suppresses the hole density further.[23] Thus, all these factors make the necessity of heavy Mg doping in order to overcome the excess electron density due to unintentional donors as well as low hole concentration due to high acceptor ionization energy. At the same time high doping levels of Mg in GaN is found to create pyramidal defects and stacking faults as well as reduction in hole density due to self compensation and compensation due to excess $V_N$ formation.[24-26] So the optimization of Mg doping levels in GaN is also a necessary and challenging process for increasing the hole concentration. Depending on the growth method there are possibilities of multiple acceptor levels due to Mg dopant and complexes formed by it.[27,28] Photoluminescence studies on Mg doped GaN give a better understanding of these acceptors formed by Mg dopants.

In the present study, Mg dopant incorporation in GaN nanowires grown in the CVD technique via VLS process is investigated. Effect of Mg dopants at the liquid/sold interface (Au-Ga/GaN) is studied in the axial growth of GaN nanowires to find the pathway of the Mg incorporation, for the first time. Structural and spectroscopic investigations reveal the presence of Mg dopants in nanowires. Heterostructure *p-n* junction is demonstrated using the as-grown *p*-type GaN nanowires on *n*-Si substrate.

Mg doped nonpolar GaN nanowires were synthesized in the atmospheric pressure CVD technique using the VLS process. Au islands were deposited on *n*-Si(111) substrates by thermal evaporation method. These substrates were separately annealed for making the Au nanoparticles at a temperature of 900 °C for 15 min in the inert atmosphere.[11] We used Ga metal (99.999%, Alfa Aesar) as a Ga source, $NH_3$ (99.999%) as reactant gas and mixture of ultra high pure (UHP)

Ar+H$_2$ (5N) as carrier gases. Mg$_3$N$_2$ (Alfa Aesar) was used as a source for incorporating Mg in the GaN nanowires. Si substrate with Au nanoparticles was kept upstream to a Ga droplet in a high pure alumina boat which was placed into a quartz tube. The temperature of the quartz tube was slowly raised to a growth temperature of 900 $^o$C with 15 $^o$C min$^{-1}$ ramp rate. Nanowires were grown for 1 hr growth time by purging 10 sccm of NH$_3$ reactant gas and 20 sccm of Ar carrier gas and intentionally terminated the growth process by reducing the temperature of the growth zone. Concentration of the Mg dopant was controlled by changing either distance between the Mg source and the substrate the flow rate of the carrier gas. Mg dopants, for *p*-type conduction in GaN nanowires were activated by thermally annealing the as-grown samples in a separate quartz tube in the N$_2$ atmosphere at 750 $^o$C for 30 min.

Morphological features of the as-grown samples were examined by field emission scanning electron microscope (FESEM, SUPRA 55 Zeiss). X-ray photoelectron spectroscopy (XPS) studies were carried out on Mg doped GaN nanowires using an X-ray source of non-monochromatic Al $K_\alpha$ (1486.6 eV) and the binding energy values were measured by referencing with respect to the C 1*s* peak. The spectra were processed by applying Shirley type background and curve fitted with mixture of Gaussian-Lorentzian line shape. For structural studies, high resolution transmission electron microscopy (HRTEM, LIBRA 200FE Zeiss) observations were performed on nanowires which were dispersed in isopropyl alcohol and transferred to the TEM Cu grids. Electron energy loss spectroscopy (EELS) studies were carried out for identifying the presence of Ga, N and Mg in a single nanowire using in-column second order corrected omega energy filter type spectrometer with an energy resolution of 0.7 eV. The distribution of the three elements Ga, N and Mg in nanowires was studied by generating energy-filtered transmission electron micrographs (EFTEM) corresponding to the core-loss energy of each element of the nanowires. The photoluminescence (PL) studies were carried out on nanowires with an excitation wavelength of 325 nm of the He-Cd laser. The spectra were collected using 2400 lines mm$^{-1}$ grating and thermoelectrically cooled CCD detector. I-V characteristics of *p*-type GaN nanowires/*n*-Si heterostructure device were studied by depositing the Au (100 nm)/Ni (25 nm) electrodes on *p*-GaN nanowires and back contacting the Si substrate with silver paste. The photoresponse of the device was tested in the reverse bias configuration at 470 nm, 530 nm and 788 nm wavelengths.

Morphological variations in the Mg doped GaN nanowires grown with two different concentrations of Mg were compared with that of the undoped nanowires. Figure 1 shows the typical FESEM micrographs of the three samples of GaN nanowires in which one is undoped (Figure 1a) and two are Mg doped, GaN:Mg-I (Figure 1b) and GaN:Mg-II (Figure 1c). The GaN:Mg-I nanowires are less doped compared to that of the GaN:Mg-II. The particle at the tip shows that the nanowires were grown in VLS process (insets in Figures 1a-1c). The diameter of the nanowires exactly follows the size of the catalyst particle at the tip and having uniform shape and size distribution of ~ 60 ($\pm$5) nm for the undoped sample (inset in Figure 1a). The growth rate of the nanowires was found to be 3 μm hr$^{-1}$. Noticeable change in the morphology and increase in the diameter distribution (65 $\pm$5 nm) was observed for the samples GaN:Mg-I compared to that of undoped nanowires. Whereas a distinct tapering at the tip region and increase in the diameter distribution to 70 ($\pm$10) nm has been observed for the GaN:Mg-II nanowires while increasing the doping concentration. The increase in diameter in the doped nanowires may be due to the increased contact angle of the catalyst in the presence of impurity.[9]

Typical XPS spectra are shown (Figure 2) for undoped and Mg doped GaN:Mg-I and GaN:Mg-II nanowires. Binding energy of Ga 3*d* level (Figure 2a) for the GaN:Mg-I nanowires was observed at 19.8 eV while for the sample GaN:Mg-II, it was observed at 19.4 eV.[29] The binding energy of N 1$s$ core level (Figure 2b) has been identified by deconvoluting the corresponding spectra from the Ga Auger electron spectra (Ga AES). N 1$s$ core level was identified at 397.5 eV (Figure 2b) for the GaN:Mg-I nanowires while it was observed at 397.1 eV for the GaN:Mg-II.[29] Binding energies of both Ga 3$d$ and N 1$s$ levels are red shifted by 0.4 eV while increasing the Mg concentration. For Mg 1$s$ core level, spectra were recorded in the binding energy range of 1320 to 1290 eV. A broad peak, observed in the energy range of 1312 to 1300 eV for GaN:Mg-I sample, is difficult to be deconvoluted for the Mg 1$s$ and Ga 2$s$ peaks (Figure 2c). It is because of the reason that both the peaks are in the same region and FWHM of the peak is quite high (~10 eV).[30] Whereas in case of the sample GaN:Mg-II, the intensity of the broad peak is increased significantly (Figure 2c). When the spectra were compared with pure Ga 2$s$ peak of the undoped sample (Figure 2c) both the shape and FWHM of the peaks were significantly different from each other. This shows the presence of Mg in doped nanowires. Since the Ga 2$s$ peak is quite asymmetric in shape, it was not possible to subtract it from the doped sample in order to estimate the Mg concentration.[30] The presence of Mg dopants in GaN nanowires was further verified in our EELS studies.

A typical low magnification TEM micrograph of an undoped nanowire shows a perfect rod like shape with uniform surface morphology (Figure 3a). It also shows an Au catalyst particle at the tip. Selected area electron diffraction (SAED) pattern (Figure 3a) reveals that the nanowire is a single crystalline wurtzite phase of GaN with zone axes along [0001]. A sharp interface between GaN nanowire and Au catalyst particle is seen in the HRTEM image (Figure 3a). An interplanar spacing of 0.275 nm (zoomed view in the inset Figure 3a) corresponds to the nonpolar {10$\bar{1}$0} planes of wurtzite GaN.[11] The growth direction of the nanowire is found to be along the [10$\bar{1}$0] direction. SAED pattern (Figure 3b) of the GaN:Mg-I nanowire, collected along the zone axis [01$\bar{1}$0] is also found to be single crystalline wurtzite phase. From the magnified view (inset Figure 3b) of the nanowire near the tip region, it is observed that the there is a distinct contrast along a small distorted portion (~20 nm in length) of the wire right bellow the catalyst particle. Sharp interface between the nanowire and Au catalyst particle has been also lost. Lattice resolved high resolution image (Figure 3b) shows an interface formed between nanowire and distorted portion of the wire where the gradual orientation of the lattice fringes of {10$\bar{1}$0} planes is observed. Variation in contrast along the wire between the catalyst particle and the interface (Figure 3b) might be due to different grain thickness formed by the excess incorporation of the Mg. Sample GaN:Mg-II, with more doping concentration than that of GaN:Mg-I, was also found to have similar distorted growth in a small portion of the wire (~40 nm length) near the tip region right below the catalyst particle (Figure 3c). An increase in diameter of this small portion is also observed, as compared to that of the remaining portion of the nanowire. SAED pattern on the nanowire collected along the zone axis [01$\bar{1}$0] shows single crystalline wurtzite phase of GaN. An interplanar spacing of 0.276 nm (Figure 3c) corresponds to the nonpolar {10$\bar{1}$0} planes of wurtzite GaN. The growth direction is found to be along the [10$\bar{1}$0].

These observations suggest that during the VLS growth process of GaN nanowire in the CVD technique, the Mg dopants may have also incorporated at liquid/solid interface of Au-Ga/GaN along with the N atoms. Incorporation of the N is expected directly at the interface, as being a group V element it cannot be dissolved in the Au catalyst.[18] On the other hand, Mg can easily be dissolved in the Au and can be infused through Au-Ga/GaN interface.[18] When the growth process is terminated by reducing the growth temperature, the excess incorporation of the Mg dopants at the tip region below the catalyst particle may have introduced stacking faults along the nanowire leading to the twining of the $\{10\bar{1}0\}$ planes.[25] However, a clear interface is observed for GaN:Mg-II nanowires at the junction of the nanowire and the distorted portion similar to that of GaN:Mg-I nanowires as the growth continues (Figures 3b and 3c).

We have carried out EELS studies to detect the presence and distribution of Mg atoms along the nanowires. A typical low intensity Mg-$K$ edge spectra (Figures 4a and 4b) collected from single GaN:Mg-I and GaN:Mg-II nanowires are presented. No appreciable changes in the shape of N-$K$ edge spectra (Figure 4c) of Mg doped nanowires were observed with respect to that of the undoped nanowires. As the extended feature of the Ga-$L$ edge is overlapping with the Mg-$K$ edge, we compare the same region with that of the undoped nanowires (Figures S1a and S1b). A significant variation in the energy loss near-edge spectroscopy (ELNES), similar to the Mg-$K$ edge features of reference MgO sample (Figures S1b) was observed for doped nanowires confirming the presence of Mg atoms in our doped samples. Distribution of Mg dopants along each nanowire of both the Mg doped samples was investigated using EFTEM micrographs generated from the electron energy loss due to Ga, N and Mg atoms. Figures 4d and 4e (grey colour) depict the typical zero-loss EFTEM micrographs of GaN:Mg-I and GaN:Mg-II nanowires, respectively. EFTEM micrographs corresponding to the Ga-$L$ edge for the two samples (red colour in Figures 4d and 4e) reveal uniform distribution of the Ga along the wire including the Au catalyst particle. This observation reveals that incorporation of the Ga has been occurred through the Au catalyst and formed the Au-Ga alloy. Similar observations were made in case of undoped GaN nanowires (Figure S2a). Whereas EFTEM micrographs for the N-$K$ edge, of both Mg doped as well as undoped nanowires (green colour in Figures 4d, 4e and supplementary Figure S2a) show uniform distribution of the N atoms along the wire except in the Au particle. The observation confirms that the N has been incorporated through the interface between Au-Ga/GaN. A faint contrast around the catalyst particle in case of the sample GaN:Mg-II (green colour in Figure 4e) is due to the small distorted portion of GaN nanowire surrounded bellow the particle, which is projected out of the plane. EFTEM micrographs corresponding to the Mg-*K* (yellow colour in Figures 4d and 4e) depict that the Mg is uniformly distributed along the nanowire up to the catalyst particle while it is absent in the catalyst particle. Since the Mg-*K* edge and extended feature of Ga-*L* edge overlap with each other, we have selected a narrow energy window at Mg-*K* edge region to generate the EFTEM micrographs with minimum effect of Ga-*L* edge and compared with that of the undoped nanowire (Figure S2b). Catalyst Au nanoparticle appears dark due to the absence of Mg in the doped sample. Whereas in case of undoped nanowire, a very faint uniform contrast was observed for the nanowire including the catalyst nanoparticles due to the extended features of Ga-*L* edge. This happened because the Ga formed alloy with Au and diffused towards the nanowire. So Au particle does not disappear in the EFTEM micrograph of the undoped nanowire. The uniform contrast of the Mg element along the nanowire with uniform shape revealed that the Mg dopants, preferably, were incorporated though the Au-Ga/GaN interface during the axial growth process. It was different for nanowires synthesized in the radial growth process using epitaxial techniques where the Mg incorporation takes place along the side facets showing accumulation of dopants along the side walls forming a shell.[18] The uniform distribution of the dopant indicates insignificant surface accumulation of Mg in the present study using the CVD growth technique. It is also clear from the micrographs that the amount of the Mg dopants is less in GaN:Mg-I (yellow colour in Figure 4d) than that of the GaN:Mg-II (yellow colour in Figure 4e) nanowires which is consistent with that of the doping levels used during growth of the samples.

Photoluminescence (PL) spectra of undoped and Mg doped GaN nanowires were studied by exciting the nanowires using an excitation source of 325 nm UV laser. The luminescence from the undoped GaN nanowires at room temperature (300 K) (Figure 5a) shows a broad band ranging from 3.3 eV to 3.5 eV. When the nanowires are cooled to 80 K, the spectra are dominated by a peak at 3.516 eV followed by the peaks at 3.475 eV and 3.33 eV (Figure 5a). The PL peak at 3.516 eV corresponds to free exciton (FE) transition between conduction band minimum and valence band maximum along with its phonon replica (FE-2LO) at 3.33 eV.[11] An intense luminescence of FE emission represents the high optical quality of the undoped nanowires. The luminescence peak at 3.475 eV (Figure 5a) is identified as the emission due to the excitons bound to neutral donors (DBE), which may appear because of the presence of either the intrinsic $V_N$ or the residual O donors formed below the conduction band.[23] Mg doped GaN:Mg-I and GaN:Mg-II nanowires (Figures 5b and 5c) show emission of a broad band around 3.2 eV to 3.3 eV, generally termed as the ultraviolet luminescence (UVL) along with low intense peaks around 2.85 eV, termed as the blue luminescence (BL)[31] and 3.46 eV at 300 K. The UVL and the BL, which are assigned to the donor acceptor pair (DAP) transitions involving Mg acceptors, are the main signatures of Mg doped GaN samples.[32] Another low intense peak at 3.46 eV corresponds to the transitions involving the acceptor bound excitons (ABE).[32]

At low temperature (80 K), PL spectra of both the Mg doped samples (GaN:Mg-I and GaN:Mg-II) are dominated by DAP peaks. The BL peak appears because of the transitions between shallow acceptors ($Mg_{Ga}$) and deep donors, presumably owing to the $V_N$. The intensity of the BL band increases with increasing Mg doping in GaN. The significant increase in the intensity of the BL band around 2.85 eV (Figures 5b and 5c) is clearly observed for our Mg doped GaN nanowires. As the doping concentration in GaN:Mg-II nanowires is higher than that for the GaN:Mg-I samples, the intensity of the BL band of the former is quite high compared to that of the later. The UVL peaks due to the DAP transitions are another significant feature in the PL spectral of the Mg doped GaN nanowires. An intense peak at 3.18 eV in the PL spectra (Figure 5b) of sample GaN:Mg-I corresponds to DAP transition (DAP1) involving shallow acceptors and shallow donors ($V_N$). The shallow acceptor is attributed to the acceptor formed by Mg as a substituent for Ga ($Mg_{Ga}$).[31] Whereas, in case of GaN:Mg-II nanowires, luminescence due to DAP (DAP1) transitions is observed around 3.18 eV along with a shoulder peak around 3.28 eV which may be due to another type of DAP transition (DAP2).[27] The observation of DAP2 transitions may arise because of two different types of acceptors having different binding energies related with Mg or due to two types of donor states related to $V_N$ and their complexes with H.[27,28] The DAP2 transition may be observed at higher temperature because of the defective nature of the band. In both the samples of Mg doped GaN nanowires, the PL peak corresponding to the ABE is suppressed by DAP peaks at 80 K. A tiny peak around 3.72 eV in the PL spectra of all the three samples (Figures 5a, 5b and 5c) corresponds to symmetry allowed Raman mode of $A_1(LO)$.[11] At 4 K, only one kind of DAP transition (DAP1) in PL spectra (Figure 5d) was observed for both the samples GaN:Mg-I and GaN:Mg-II revealing unstable nature of DAP2.[27] The peak position of the DAP1 is red shifted by 20 meV for the sample GaN:Mg-II with increasing the Mg doping level than that of the GaN:Mg-I nanowires. During the growth process H evolved from $NH_3$ and carrier gases passivated both the acceptors ($Mg_{Ga}$) and donors ($V_N$) by forming the complexes of $Mg_{Ga}$-H and $V_N$-H, respectively.[5] Presence of these two complexes may manifests in the observation of DAP2 peak, which is observed to be generally absent in our low temperature PL studies. During the post-growth annealing process both the complexes dissociate and activate the $Mg_{Ga}$.[23] Thus, in the absence of DAP2 in the low temperature PL studies show the successful incorporation and activation of different concentrations of Mg dopants in the nonpolar GaN nanowires.

I-V characteristics of the Mg doped *p*-type nonpolar GaN nanowires on *n*-Si substrate show a *p-n* junction diode behaviour under dark (Figure 6). In the forward bias, the turn-on voltage of the device is found to be 0.7 V and the ratio of the forward bias to the reverse bias current under dark is found to be $10^3$ at ±1 V. The *p-n* junction behaviour for the heterojunction of Mg doped GaN nanowires on *n*-Si(111) substrates, thus confirms the presence of activated Mg atoms as *p*-type dopants.[3] I-V characteristics of the undoped-GaN/*n*-Si heterojunction under dark shows typical Schottky behaviour (Figure S3). The photo response of the *p-n* junction was tested by illuminating the device with 470 nm, 530 nm and 788 nm wavelengths at ±4 V. The change in the current was observed for all the wavelengths only under reverse bias configuration of the *p-n* heterojunction. The on-off ratios of the photocurrent in the reverse bias mode for all the three wavelengths are found to be in the order of $10^2$. The photocurrent is most probably from the *n*-region (Si) of the *p-n* diode, as the excitation energies are quite below the band gap of GaN (~ 3.4 eV at 300 K).

**Summary**

Mg doped *p*-type nonpolar GaN nanowires are successfully grown via vapour-liquid-solid mechanism in the chemical vapor deposition technique. High resolution transmission electron microscopic (HRTEM) studies show the nanowires are grown along [10$\bar{1}$0] nonpolar planes of GaN. X-ray photoelectron and electron energy loss spectroscopy studies on Mg doped nanowires show the presence of Mg dopants. Local distortion and twining of atomic planes between catalyst particle and the nanowire confirm the interface as Mg incorporation pathway. Structural analysis shows the direct evidence of incorporation of the Mg atoms through the Au-Ga/GaN interface for the first time using the energy filtered TEM imaging. Photoluminescence studies with DAP transition indicate successful activation of Mg dopants. The *p-n* junction behaviour of *p*-GaN/*n*-Si(111) heterojunction under dark confirm the activation of Mg atoms as *p*-dopants in GaN nanowires. A very high on-off ratio of ~$10^2$, observed for the photoresponse of these *p-n* heterojunction, can be used in possible photodetector applications.

## ACKNOWLEDGMENT

One of us (AP) acknowledges Department of Atomic Energy for the financial aid. We thank Pallabi Pramanik and Sayantani Sen of Institute of Radiophysics and Electronics, University of Calcutta for their help in 4 K low temperature photoluminescence studies, A. K. Sivadasan of MSG, IGCAR for his help in experiments and Arindam Das, MSG, IGCAR for his useful discussions. We acknowledge A. K. Tyagi of Materials Science Group, IGCAR for his encouragement and support. We would like to acknowledge the Center for Research in Nanoscience and Nanotechnology, Kolakata for facilitating optical studies.

## SUPPORTING INFORMATION

EELS, EFTEM and I-V characteristics of udoped GaN nanowire are discussed with necessary figures. This information is available free of charge via the Internet at http://pubs.acs.org

**Figure Captions**

**Figure 1.** Typical FESEM images of GaN nanowires. (a) Undoped nanowires having uniform diameter along with the Au catalyst particle (encircled) at the tip of the wire (inset). (b) GaN:Mg-I and (c) GaN:Mg-II are Mg doped GaN nanowires. The sample GaN:Mg-I is less doped compared to that of the sample GaN:Mg-II. All the nanowires are grown with Au catalyst particle (encircled) at the tip (insets Figures b and c).

**Figure 2.** XPS spectra from Mg doped GaN nanowires of the samples GaN:Mg-I and GaN:Mg-II. (a) Ga 3$d$ (b) N 1$s$ and (c) Ga 2$s$ + Mg 1$s$. Ga 2$s$ peak recorded from undoped sample is plotted in (c) for comparison. Solid lines are experimental data and discontinuous lines are fitted data.

**Figure 3.** (a) TEM low magnification image of a typical undoped GaN nanowire. SAED of the nanowire, indexed to wurtzite GaN with zone axis along [0001]. HRTEM collected near the tip of the nanowire having growth direction along [10$\bar{1}$0] and a sharp interface between Au and GaN is seen (inset). (b) GaN nanowire doped with Mg (GaN:Mg-I). SAED of the nanowire collected along the zone axis [01$\bar{1}$0] of wurtzite phase. An interface separating the nanowire and distorted portion of the tip region with Au nanoparticle is seen in HRTEM. Inset, shows the contrast variation in the tip region. (c) GaN nanowire doped with increased Mg concentration (GaN:Mg-II). SAED of the nanowire collected along the zone axis [01$\bar{1}$0] of wurtzite phase. Clear interface is visible in the HRTEM collected near the tip region and nanowire having growth direction along [10$\bar{1}$0]. Inset, shows lattice fringes of (10$\bar{1}$0) planes at the interface.

**Figure 4.** Typical EELS Mg-$K$ edge spectra collected from single GaN nanowires of the each Mg doped sample (a) GaN:Mg-I and (b) GaN:Mg-II. (c) N-$K$ edge spectra collected from the nanowires of three samples. EFTEM micrographs generated from the nanowire for the each Mg doped sample (d) GaN:Mg-I and (e) GaN:Mg-II correspond to zero-loss energy (grey colour), Ga-*L* (red colour), N-*K* (green colour) and Mg-*K* (yellow colour). These are pseudo colors chosen for describing presence of different elemental species in the nanowire.

**Figure 5.** Typical photoluminescence spectra of GaN nanowires of the samples (a) undoped, (b) and (c) Mg doped in increased concentration, GaN:Mg-I and GaN:Mg-II respectively, collected at the temperatures 300 K and 80 K. (d) DAP1 peak at 4 K for the nanowires of Mg doped samples GaN:Mg-I and GaN:Mg-II. Discontinuous lines are fitted data corresponding to the individual peaks.

**Figure 6.** I-V characteristics of the *p*-GaN/*n*-Si heterojunction under dark (circle) and under the illumination of 470 nm, 530 nm and 788 nm wavelengths.

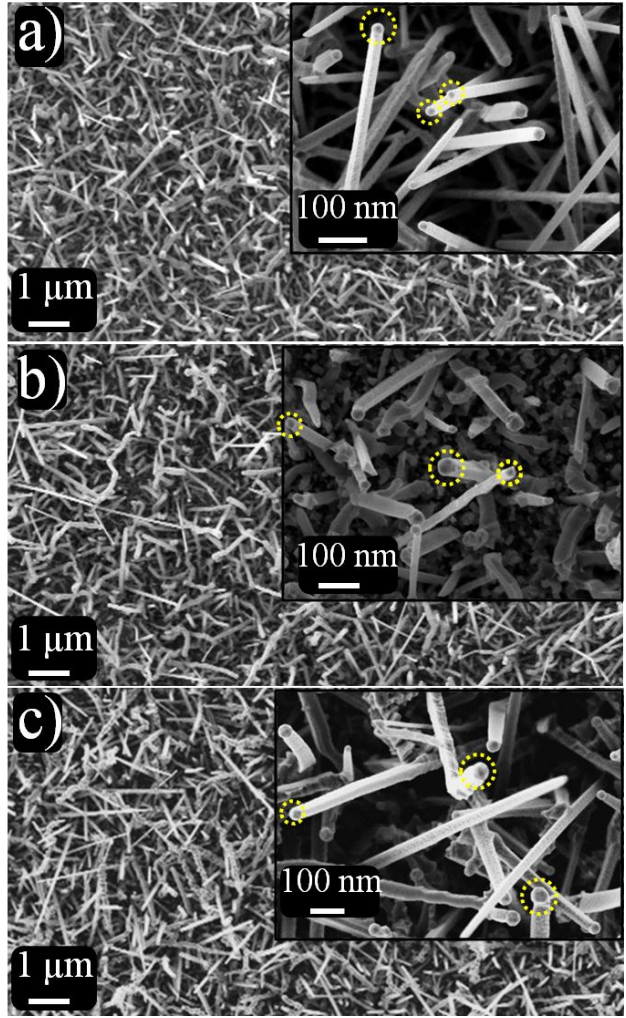

**Figure 1.** Typical FESEM images of GaN nanowires. (a) Undoped nanowires having uniform diameter along with the Au catalyst particle (encircled) at the tip of the wire (inset). (b) GaN:Mg-I and (c) GaN:Mg-II are Mg doped GaN nanowires. The sample GaN:Mg-I is less doped compared to that of the sample GaN:Mg-II. All the nanowires are grown with Au catalyst particle (encircled) at the tip (insets Figures b and c).

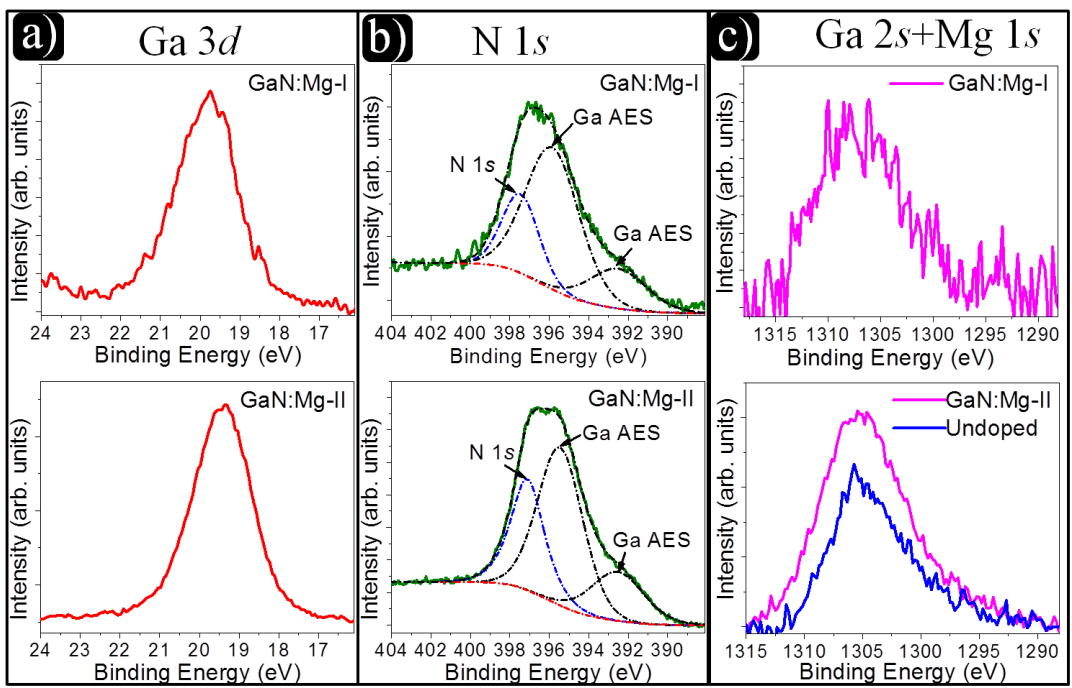

**Figure 2.** XPS spectra from Mg doped GaN nanowires of the samples GaN:Mg-I and GaN:Mg-II. (a) Ga 3*d* (b) N 1*s* and (c) Ga 2*s* + Mg 1*s*. Ga 2*s* peak recorded from undoped sample is plotted in (c) for comparison. Solid lines are experimental data and discontinuous lines are fitted data.

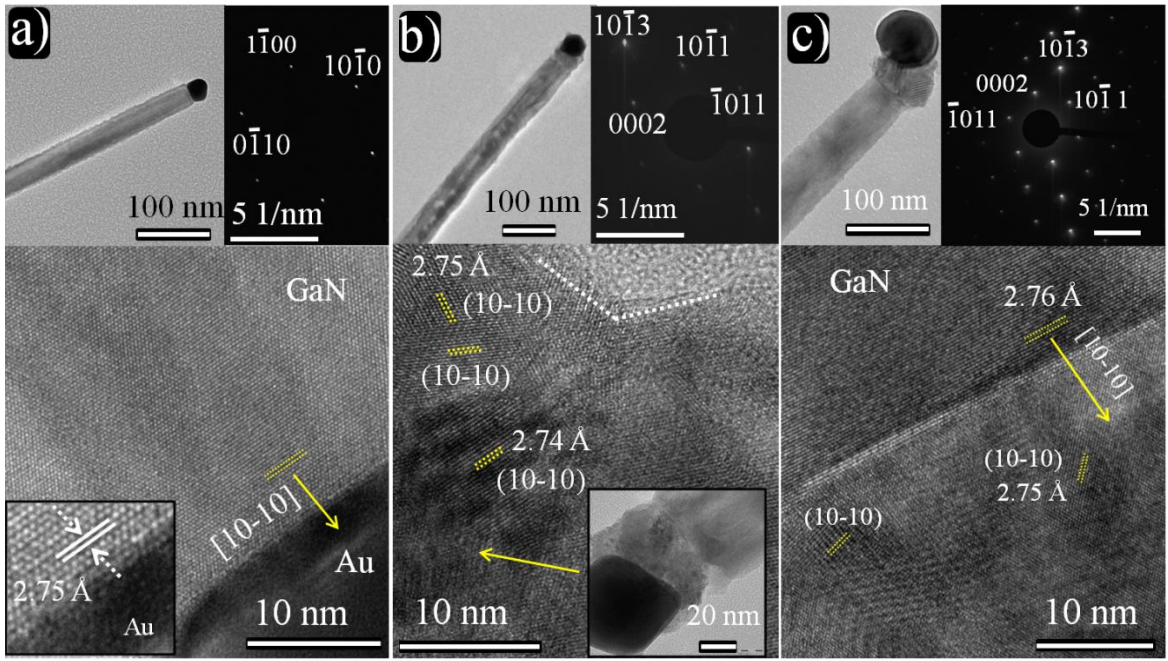

**Figure 3.** (a) TEM low magnification image of a typical undoped GaN nanowire. SAED of the nanowire, indexed to wurtzite GaN with zone axis along [0001]. HRTEM collected near the tip of the nanowire having growth direction along [10$\bar{1}$0] and a sharp interface between Au and GaN is seen (inset). (b) GaN nanowire doped with Mg (GaN:Mg-I). SAED of the nanowire collected along the zone axis [01$\bar{1}$0] of wurtzite phase. An interface separating the nanowire and distorted portion of the tip region with Au nanoparticle is seen in HRTEM. Inset, shows the contrast variation in the tip region. (c) GaN nanowire doped with increased Mg concentration (GaN:Mg-II). SAED of the nanowire collected along the zone axis [01$\bar{1}$0] of wurtzite phase. Clear interface is visible in the HRTEM collected near the tip region and nanowire having growth direction along [10$\bar{1}$0]. Inset, shows lattice fringes of (10$\bar{1}$0) planes at the interface.

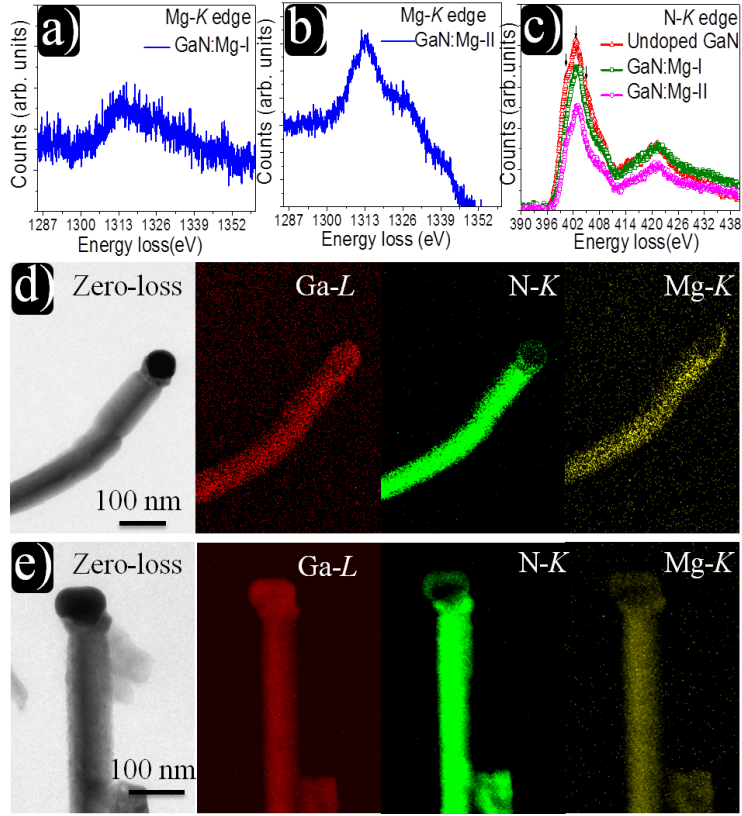

**Figure 4.** Typical EELS Mg-*K* edge spectra collected from single GaN nanowires of the each Mg doped sample (a) GaN:Mg-I and (b) GaN:Mg-II. (c) N-*K* edge spectra collected from the nanowires of three samples. EFTEM micrographs generated from the nanowire for the each Mg doped sample (d) GaN:Mg-I and (e) GaN:Mg-II correspond to zero-loss energy (grey colour), Ga-*L* (red colour), N-*K* (green colour) and Mg-*K* (yellow colour). These are pseudo colors chosen for describing presence of different elemental species in the nanowire.

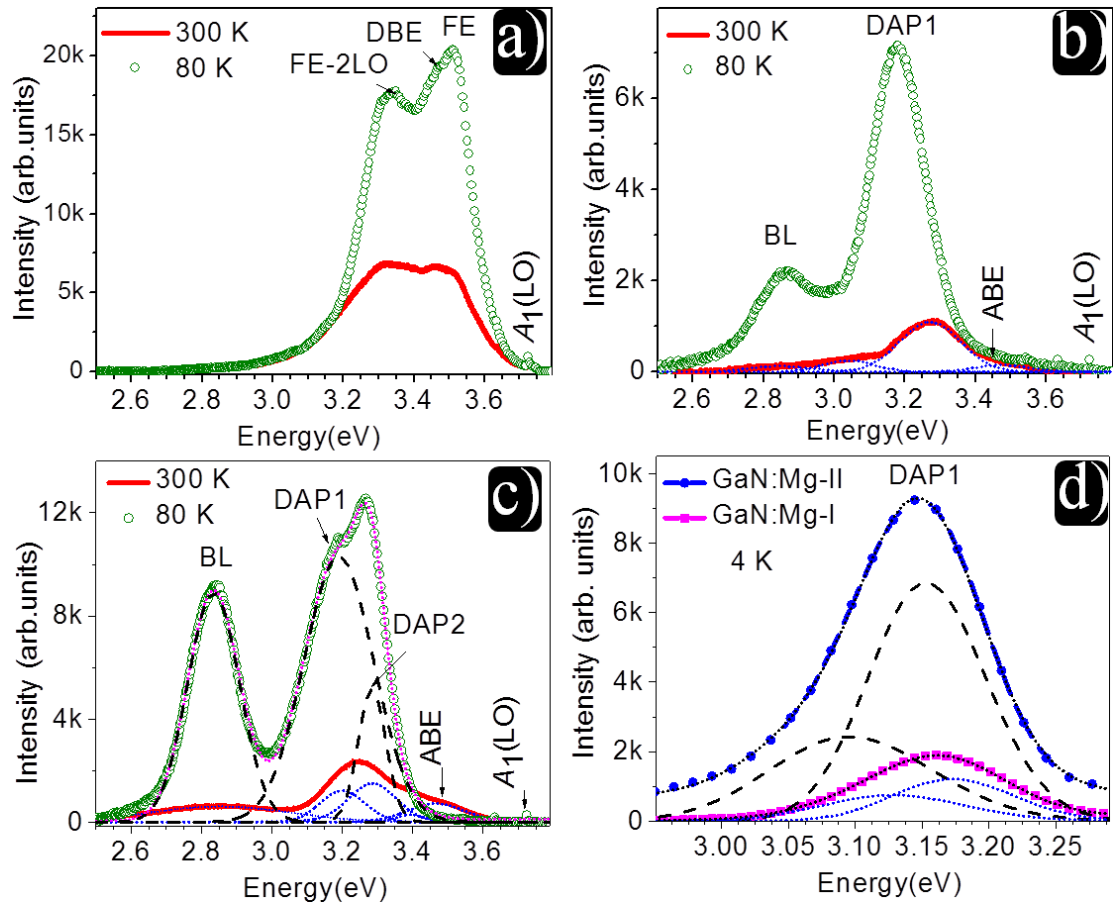

**Figure 5.** Typical photoluminescence spectra of GaN nanowires of the samples (a) undoped, (b) and (c) Mg doped in increased concentration, GaN:Mg-I and GaN:Mg-II respectively, collected at the temperatures 300 K and 80 K. (d) DAP1 peak at 4 K for the nanowires of Mg doped samples GaN:Mg-I and GaN:Mg-II. Discontinuous lines are fitted data corresponding to the individual peaks.

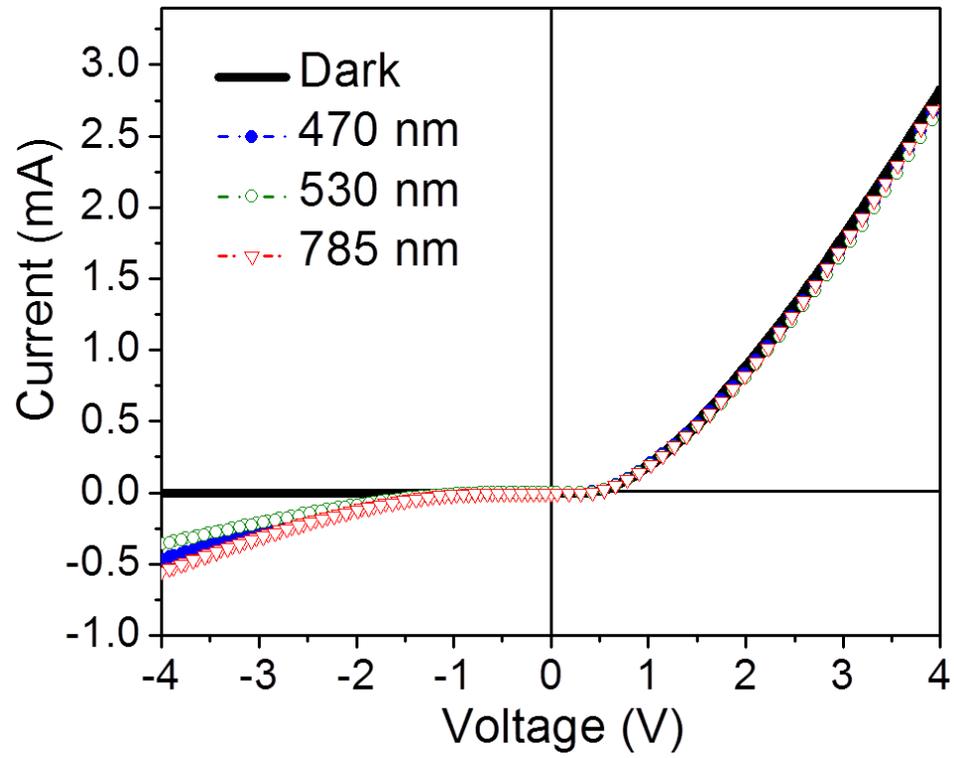

**Figure 6.** I-V characteristics of the *p*-GaN/*n*-Si heterojunction under dark (circle) and under the illumination of 470 nm, 530 nm and 788 nm wavelengths.

**Supporting information:**

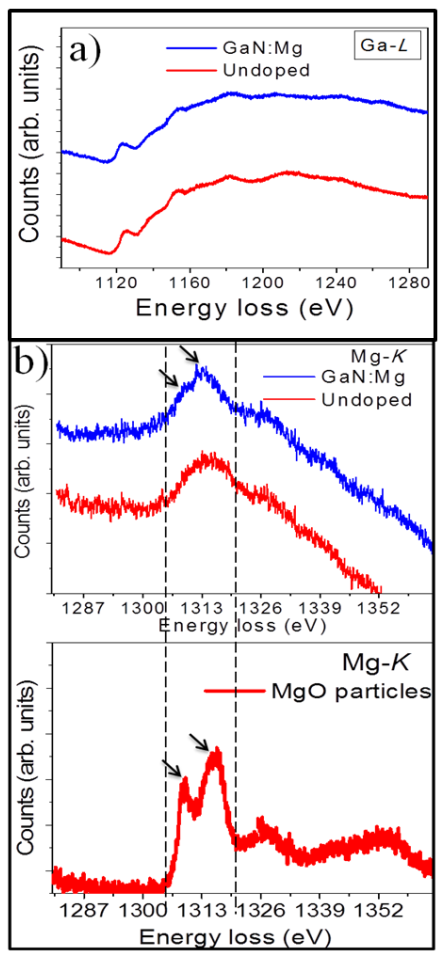

**Figure S1.** Comparison of EELS spectra corresponding to (a) Ga-*L* edge and (b) Mg-*K* edge collected from single GaN nanowires of the undoped and Mg doped samples. The data is also compared with MgO reference sample (bottom) showing similar fine feature as observed in the Mg doped GaN nanowires.

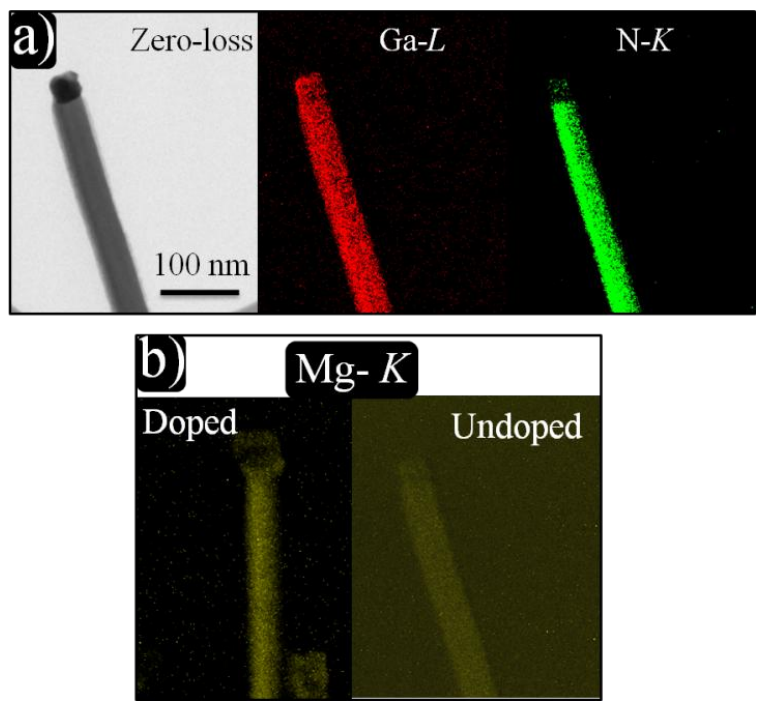

**Figure S2.** (a) EFTEM micrographs collected from the undoped nanowire (b) Comparison of the EFTEM micrographs generated from the Mg-*K* edge region, collected from doped and undoped nanowires. Au catalyst particle appears dark in the doped nanowire in the absence of Mg at the tip. Whereas in case of undoped sample, uniform contrast is observed including the catalyst particle due to the extended features of Ga-*L* edge.

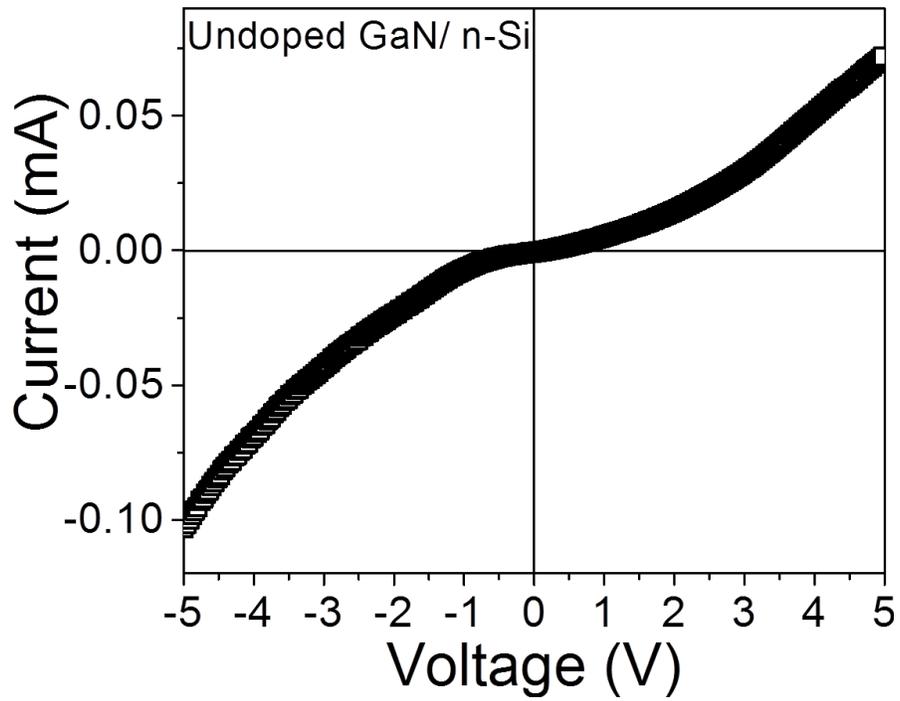

**Figure S3.** I-V characteristics of the undoped-GaN/*n*-Si heterojunction under darks showing Schottky nature.